\documentclass[aps,prl,twocolumn,superscriptaddress,showpacs,amsmath]{revtex4}

\usepackage{graphicx,amssymb}

\begin{document}

%

\title{Quantum-magneto oscillations in a supramolecular Mn(II)-[3~$\times$~3] grid}

\author{O. Waldmann}
\email[Corresponding author.\\E-mail: ]{waldmann@mps.ohio-state.edu} \affiliation{Department of Physics, The
Ohio State University, Columbus, OH 43210, USA}

\author{S. Carretta}
\author{P. Santini}
\affiliation{INFM and Dipartimento di Fisica, Universit\`a di Parma, I-43100 Parma, Italy}

\author{R. Koch}
\affiliation{Physikalisches Institut III, Universit\"at Erlangen-N\"urnberg, 91058 Erlangen, Germany}

\author{A. G. M. Jansen}
\affiliation{CEA-Grenoble, DRFMC/SPSMS, 38054 Grenoble, France} \affiliation{Grenoble High Magnetic Field
Laboratory, CNRS and MPI-FKF, 38042 Grenoble, France}

\author{G. Amoretti}
\affiliation{INFM and Dipartimento di Fisica, Universit\`a di Parma, I-43100 Parma, Italy}

\author{R. Caciuffo}
\affiliation{INFM and Diprtimento di Fisica, Universit\`a Politecnica della Marche, I-60131 Ancona, Italy}

\author{L. Zhao}
\author{L. K. Thompson}
\affiliation{Department of Chemistry, Memorial University, St. John's, Newfoundland, Canada A1B 3X7}

\date{\today}

\begin{abstract}
The magnetic grid molecule Mn(II)-[3~$\times$~3] has been studied by high-field torque magnetometry
at $^3$He temperatures. At fields above 5~T, the torque vs. field curves exhibit an unprecedented
oscillatory behavior. A model is proposed which describes these magneto oscillations well.
\end{abstract}


\maketitle

%

Among the known magneto-oscillatory effects, the de Haas-van Alphen (dHvA) effect in metals (and the group of
effects related to it) is the most prominent example, having had a formative influence on our modern picture
of solid state physics \cite{Sho84}. The dHvA effect originates from a quantization of closed electron orbits
into Landau levels in a magnetic field, so that the density of states at the Fermi energy exhibits a markedly
oscillatory evolution as a function of magnetic field. Currently, it finds application in a wide range of
materials, e.g. heavy fermion compounds \cite{UPT3}, low-dimensional organic metals \cite{OrgMet},
high-temperature superconductors \cite{HiTc}, two-dimensional electron gases \cite{2DEG}, or in the recently
discovered superconductor MgB$_2$ \cite{MgB2}.

In this work, we report on a new magneto-oscillatory effect, discovered in the molecular nanomagnet
[Mn$_9$(2POAP-2H)$_6$](ClO$_4$)$_6$ $\cdot$ 3.57 MeCN $\cdot$ H$_2$O, the so called Mn-[3~$\times$~3] grid.
Molecular nanomagnets are compounds with many magnetic metal ions linked by organic ligands to form well
defined magnetic nanoclusters. They have attracted much interest since they can exhibit fascinating quantum
effects at the mesoscopic scale. For instance, quantum tunneling of the magnetization has been observed in
the clusters Mn$_{12}$ or Fe$_8$ \cite{Mn12_Fe8}. We have performed high-field torque magnetometry on the
Mn-[3~$\times$~3] grids at $^3$He temperatures and observed striking magneto oscillations in the torque
signal. We show that they arise from the interplay between antiferromagnetic interactions within a molecule
and Zeeman splitting on the one hand and the magnetic anisotropy on the other hand.

\begin{figure}[b]
\includegraphics{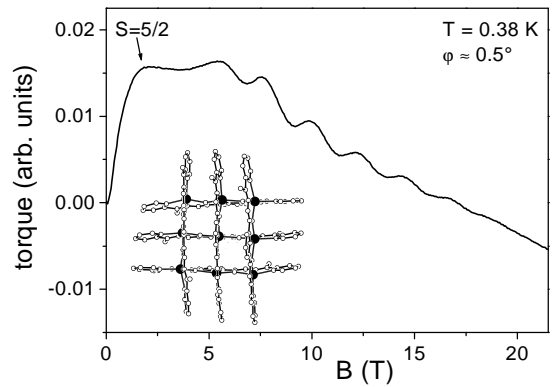}
\caption{\label{fig1}Torque vs. magnetic field of a Mn-[3~$\times$~3] single crystal at low temperature. As
the magnetic field lies almost in the plane of the grid molecule, the signal is very small and might be
affected by a significant, but smooth background (e.g. due to Faraday forces). The peak at $\approx$ 2~T
stems from the zero-field splitting of the $S = 5/2$ ground state. The inset shows the structure of the
cation [Mn$_9$(2POAP-2H)$_6$]$^{6+}$ (with Mn in black and H atoms omitted).}
\end{figure}

In the Mn-[3~$\times$~3] grid molecules, nine spin-5/2 Mn(II) ions occupy the positions of a regular
3~$\times$~3 matrix, held in place by a lattice of organic ligands [inset of Fig.~\ref{fig1}]. These grids
were characterized recently by magnetization and torque measurements and found to exhibit unusual magnetic
properties \cite{OW_Mn3x3}. The Mn ions within a molecule experience an antiferromagnetic interaction leading
to a total spin $S=5/2$ ground state at zero field. At a field of about 7~T, the ground state changes
abruptly to a new state, a $S=7/2$ level, accompanied by a change of the sign of the magnetic anisotropy.
Notably, intermolecular magnetic interactions are at best on the order of few 10~mK. This is evident from the
crystal structure (the smallest separation between two molecules in the crystal is $>$ 8~\AA), but has been
checked also experimentally \cite{insulator}. Accordingly, even at $^3$He temperatures the magnetism of a
macroscopic crystal sample reflects that of a single molecule.

%

Single crystals of Mn-[3~$\times$~3] were prepared as reported \cite{Zha00}. They crystallize in the space
group $C_2/c$. The cation [Mn$_9$(2POAP-2H)$_6$]$^{6+}$ exhibits a slightly distorted $S_4$ molecular
symmetry, the $C_2$ axis is perpendicular to the grid plane. The average distance between the Mn(II) ions is
3.93~\AA, the smallest distance between clusters is larger than 8~\AA. Consistent with the planar structure,
the Mn-[3~$\times$~3] grid exhibits a practically uniaxial magnetic behavior, and uniaxial and molecular
$C_2$ symmetry axes coincide \cite{OW_Mn3x3}. Magnetic torque was measured with a homemade silicon cantilever
device \cite{OW_Mn3x3,Koc03} inserted into the M10 magnet at the Grenoble High Magnetic Field Laboratory
using a $^3$He evaporation cryostat. Resolution of the sensor was about 10$^{-11}$~Nm, the weight of the
crystals investigated was about 20~$\mu$g, nonlinearity was less than 2\%, and the accuracy of the {\it in
situ} alignment of the $C_2$ crystal axis with respect to the magnetic field was $0.3^\circ$. Only raw torque
data are shown here. Background signals were, if not noted otherwise, negligible. Temperature was measured
with a RuO$_2$ thick film resistor.

%

\begin{figure}
\includegraphics{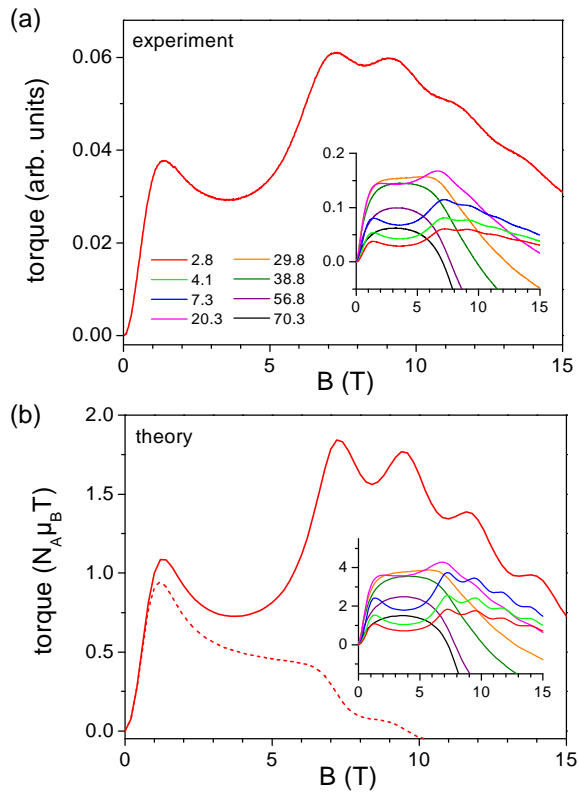}
\caption{\label{fig2}(a) Experimental and (b) calculated torque curves as function of magnetic field of a
Mn-[3~$\times$~3] single crystal at 0.4~K for various orientations of the magnetic field (the legend gives
the angle $\varphi$ between field and grid plane). The main panel shows the curves for $\varphi = 2.8^\circ$.
The dashed line in panel (b) represents the calculated torque curve for 0.4~K and 2.8$^\circ$, but with
effects of level mixing artificially forced to zero.}
\end{figure}

Figure~\ref{fig1} presents the field dependence of the torque of a Mn-[3~$\times$~3] single crystal at
0.38~K. The magnetic field was applied almost perpendicular to the $C_2$ axis of the molecule; the angle
$\varphi$ between magnetic field and plane of the grid was about $0.5^\circ$. At low fields, the torque rises
quickly to reach a maximum at about 2~T. This part of the signal stems from the $S = 5/2$ ground state and
shows the expected behavior \cite{Cor00}. At higher fields, however, above about 5~T, the overall torque
signal decreases with the superimposition of pronounced oscillations, which fade out with increasing field
strength. Such magneto oscillations have not been observed in molecular nanomagnets so far \cite{FW_osc}, and
their observation is a main result of this work.

An overview of the dependence of the torque curves on the magnetic field orientation is given in
Fig.~\ref{fig2}(a). The oscillations are clearly visible for angles smaller than about 10$^\circ$. For larger
angles, the torque decreases strongly at higher fields and actually crosses the zero line, consistent with
the observations in Ref.~[\onlinecite{OW_Mn3x3}] (the torque measurements reported there did not show the
oscillations because they were performed at too high a temperature).

%

The magnetic properties of molecular metal complexes are generally described by a spin Hamiltonian, which,
assuming an idealized structure, for Mn-[3~$\times$~3] becomes

\begin{eqnarray}
\label{H3x3}
 H &=& -J_R \left( \sum^7_{i=1}{ \textbf{S}_i \cdot \textbf{S}_{i+1} }+ \textbf{S}_8 \cdot \textbf{S}_1 \right)
\cr
 && -J_C \left( \textbf{S}_2 + \textbf{S}_4 + \textbf{S}_6 + \textbf{S}_8 \right) \cdot \textbf{S}_9
 \cr
 &&  + D_R \sum^8_{i=1} S^2_{i,z}  + D_C S^2_{9,z} + g \mu_B \textbf{S} \cdot \textbf{B}
\end{eqnarray}

with ${\bf S} = \sum_{i} {\bf S}_i$. It consists of the isotropic nearest-neighbor exchange terms, the second
order ligand-field terms and the Zeeman terms \cite{Dipdip}. $J_R$ characterizes the couplings of the eight
peripheral Mn ions, $J_C$ those involving the central Mn ion (spins at "corners" are numbered 1, 3, 5, 7;
those at "edges" 2, 4, 6, 8; and the central spin 9).

The dimension of the Hilbert space is huge for Mn-[3~$\times$~3]. However, recently it has been shown that
the field dependent low-temperature properties can be very well described by an effective spin Hamiltonian,
in which the spin operators for the corner and edge spins are combined to sublattice spin operators ${\bf
S}_A$ and ${\bf S}_B$, respectively, with $S_A = S_B = 4 \times 5/2$ \cite{OW_3x3_NVT}. Physically, this
approach works well because the internal spin structure due to the dominant Heisenberg interaction is
essentially classical \cite{OW_Cr8}. In this approach, Mn-[3~$\times$~3] is described by

\begin{eqnarray}
\label{Heff}
 H^{3\times3}_{eff} &=& - \tilde{J}_R \textbf{S}_A \cdot \textbf{S}_B
 + \tilde{D}_R ( S^2_{A,z} + S^2_{B,z} )
 \cr
 && - J_C \textbf{S}_B \cdot \textbf{S}_9  + D_C S^2_{9,z} + g \mu_B \textbf{S} \cdot \textbf{B},
\end{eqnarray}

with $\tilde{J}_R = 0.526 J_R$, $\tilde{D}_R =  0.197 D_R$ \cite{OW_3x3_NVT}. In
Ref.~[\onlinecite{OW_Mn3x3}], $|J_C| \ll |J_R|$ was suggested. With this parameter regime, however, the
experimental results could not be reproduced satisfactorily. We thus tried $J_C = J_R \equiv J$, and also set
$D_C = D_R \equiv D$, as suggested by recent inelastic neutron scattering (INS) measurements
\cite{Mn3x3_ins}. As long as $J_C$ and $J_R$ are not too different, the calculated properties were found to
be only weakly affected. A similar situation holds for $D_C$ and $D_R$. $J$ and $D$ were determined such that
the position of the first ground state level crossing and the zero-field splitting of the $S = 5/2$ ground
state in the calculated energy spectrum match the experimental values of $\approx$ 7~T and $\approx$ 3~K,
respectively \cite{OW_Mn3x3}. The result is $J$ = -5.0~K and $D$ = -0.14~K, in agreement with INS experiments
\cite{Mn3x3_ins}.

The torque curves calculated with $H^{3\times3}_{eff}$ using the given parameters are presented in
Fig.~\ref{fig2}(b). Apparently, the model Hamiltonian reproduces the data, and in particular the
oscillations, very well \cite{levelmix}. Their origin shall be discussed in the following in more detail.

\begin{figure}
\includegraphics{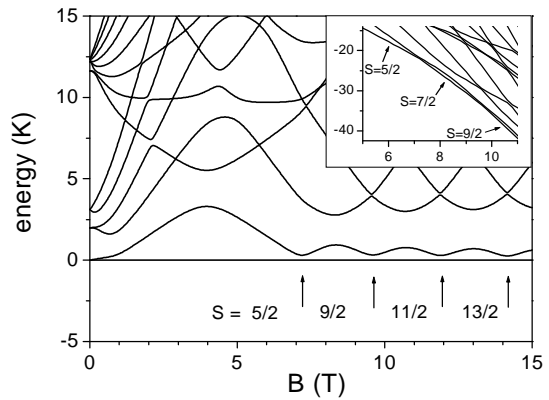}
\caption{\label{fig3}Energy spectrum vs. magnetic field for Mn-[3~$\times$~3] at $\varphi = 2.8^\circ$
(energy of lowest state was set to zero at each field). The arrows indicate the ground state level crossings,
and the numbers the (approximate) total spin quantum number $S$ of the respective ground states. The inset
details the energy spectrum near the first two level crossing, where the ground state changes as $S= 5/2
\rightarrow 7/2$ and $S= 7/2 \rightarrow 9/2$. In contrast to the main panel, energies are given here with
respect to the ground state energy at zero field.}
\end{figure}

For illustration, Fig.~\ref{fig3} shows the calculated energy spectrum as a function of magnetic field for an
angle of $\varphi = 2.8^\circ$. As magnetic anisotropy is weak, $|D| \ll |J|$, the total spin ${\bf S}$ is
almost conserved. It is convenient to rotate the spin operators such that $S_z'$ is parallel to the magnetic
field {\bf B}. Levels may then be classified by the eigenvalues of ${\bf S}^2$ and $S_z'$, $S$ and $M$, and
be written as $|S,M\rangle$.

At zero field, the dominant isotropic exchange results in a $S=5/2$ ground state followed by excited states
with $S$ = 7/2, 9/2, 11/2, $\ldots$. Upon application of a magnetic field, the levels are shifted by the
Zeeman energy, $g \mu_B M B$, leading to a series of ground state level crossings (LCs) at characteristic
fields at which the ground state switches abruptly from $5/2 \rightarrow 7/2$, $7/2 \rightarrow 9/2$, $9/2
\rightarrow 11/2$, and so on (we abbreviated $|S,-S\rangle$ by the value of $S$). This is shown for the first
two LCs in the inset of Fig.~\ref{fig3}. The change of the slope of the ground state from $-5/2 g \mu_B$ to
$-7/2 g \mu_B$ and $-9/2 g \mu_B$ is apparent.

The effect of the ligand-field terms is to mix spin levels with $|\Delta S| = 0,1,2$ \cite{Ben90,OW_MQT_FW}.
The $\Delta S=0$ mixing leads to a zero-field splitting of each spin multiplet, which, e.g., is clearly
visible in Fig.~\ref{fig3} for the six levels of the $S=5/2$ ground state. Importantly, because of $|\Delta
S|=1$, the ligand field terms also mix the levels which are involved in a LC. This results in a repulsion of
the two states whenever they come close, and the LCs split, as is seen by the energy gaps at the LC fields in
Fig.~\ref{fig3} (which are about 0.3~K at the angle of 2.8$^\circ$).

The abrupt changes of the magnetic character of the ground state at the LCs lead to a step-like field
dependence of the magnetization and torque. These magnetization and torque steps are frequently observed
\cite{Sha02}, e.g. in molecular ferric wheels \cite{FWs,Cor99,OW_FWs,FW_osc}. However, as we will show now,
the level mixing at the LCs lead also to peak-like contributions in the torque (but not the magnetization).
For the sake of completeness, we mention that mixing of levels with $|\Delta S|=2$ is pronounced only for
excited states \cite{OW_MQT_FW}, and is of little relevance here.

%

Near each LC only two states, $|S,-S\rangle$ and $|S+1,-S-1\rangle$, are relevant at low temperatures, and
the energy spectrum can be described by a two-level Hamiltonian $H_{S,S+1} =
\left(\begin{array}{cc}\epsilon_S & \Delta/2 \\\Delta/2 & \epsilon_{S+1}\end{array}\right)$. Here,
$\epsilon_S(B,\varphi)$ and $\epsilon_{S+1}(B,\varphi)$ describe the field and angle dependence of the two
levels {\it excluding} level mixing at the LC, but {\it including} the effects of the zero-field splitting.
The level mixing itself is parameterized by a possibly field and angle dependent $\Delta$. Up to first order,
the energies $\epsilon_S$ can be written as $\epsilon_S(B,\varphi) = - b S + \Delta_S(\varphi)$ with the
reduced field $b = g \mu_B B$ \cite{Cor99,OW_FWs}. $\Delta_S(\varphi)$ accounts for the magnetic interaction
and the zero-field splitting and thus does not depend on magnetic field. The LC field is given by
$b_c(\varphi)= \Delta_{S+1} - \Delta_S$. Importantly, in first order the level mixing is also independent of
the field: It is determined by the matrix element between the two states and the ligand-field terms in
Eq.~\ref{H3x3}, $\Delta = \langle S,-S| H'_{LF} |S+1,-S-1\rangle$. Thus $\Delta = \Delta(\varphi)$.

At zero temperature, the magnetic moment is given by $m = - \partial \epsilon /
\partial B$, where $\epsilon$ is the lowest energy as calculated from $H_{S,S+1}$. We decompose
the magnetic moment into $m = m_S + \delta m$. Here, $m_S$ is the magnetic moment for fields well below the
LC field $b_c$ so that $\delta m$ describes the changes of $m$ at the LC. Similarly, the torque $\tau = -
\partial \epsilon / \partial \varphi$ is decomposed into $\tau = \tau_S + \delta \tau$. In our model, one
obtains

\begin{eqnarray}
\label{mtau}
 \delta m(b,\varphi) = {1\over 2} \left[ 1 + (b-b_c) G(b,\varphi) \right]
 \\
 \delta \tau(b,\varphi) = \delta m {\partial b_c \over \partial\varphi} - {1\over 2} G(b,\varphi)
\Delta {\partial \Delta \over \partial\varphi},
\end{eqnarray}

with $G(b,\varphi) = [ (b_c-b)^2 + \Delta^2]^{-1/2}$. Therefore, as function of magnetic field the magnetic
moment exhibits a step at the LC which is broadened by the level mixing (and by temperature if $T > 0$). In
contrast, the torque consists of two contributions. The first term, being proportional to $\delta m$, leads
to a broadened torque step. These magnetization and torque steps, as mentioned already, are well known
\cite{FWs,Cor99,OW_FWs,Sha02}. The second term, however, whose field dependence is controlled by
$G(b,\varphi)$ only, leads to a peak-like contribution centered at the LC field $b_c$. This is apparent by
noting that $G(b,\varphi)$ is basically the square root of the Lorentz function. Importantly, as this peak
arises only for nonzero level mixing it is a direct signature of it. It is absent in the magnetic moment
since the level mixing $\Delta$ is markedly angle dependent but virtually field independent.

Thus, the torque curve consists of a series of steps and peaks at each LC field. For Mn-[3~$\times$~3], the
steps are small and the peaks at each LC field superimpose to produce an oscillatory field dependence. This
is clearly seen in Fig.~\ref{fig2}(b), which also presents the 2.8$^\circ$ torque curve as it is calculated
with the level mixing at the LCs set to zero artificially (i.e., only mixing of levels with $\Delta S = 0$
was retained). This curve exhibits only thermally broadened steps at the LC fields, demonstrating the
connection of the oscillations with nonzero level mixing.

In view of the simplicity of our model Hamiltonian, the agreement between theoretical and experimental curves
is really good, though it is not perfect. Most notably, the experimental peaks are significantly broader and
smaller than the theoretical ones. The agreement can be improved by using a larger temperature value in the
calculation, $T_{eff} = T + x$. We call $x$ the Dingle temperature because of the analogy to the practice in
dHvA work \cite{Sho84}, but emphasize the different underlying physics. The experiment suggests $x \approx$
80~mK. Several intrinsic and extrinsic effects can be envisaged to be responsible for this additional
broadening and more detailed investigations are needed.

%

In conclusion, in the molecular grid Mn-[3~$\times$~3] we have observed a new type of quantum-magneto
oscillations in the field dependence of the magnetic torque. We have shown that they are associated with
level crossings which appear regularly as a function of field due to the combined effect of the magnetic
interaction and the Zeeman term. These level crossings are split because of pronounced level mixing induced
by the magnetic anisotropy which are accompanied by peak-like signals in the torque producing the
spectacular, oscillatory field dependence. This effect should be observable in a number of other molecular
magnetic clusters, like e.g. dimers of two different magnetic centers with an antiferromagnetic coupling. The
magnetic oscillations demonstrated in this study provide direct access to quantum properties, which will help
our understanding of molecular nanomagnets.

%

\begin{acknowledgments}
We thank N. Magnani, E. Liviotti and T. Guidi for enlightning discussions.
Financial support by the Deutsche Forschungsgemeinschaft, the Department of Energy (Grant No.
DE-FG02-86ER45271), and through the TMR Program of the European Community is gratefully acknowledged.
\end{acknowledgments}

%

%

\begin{references}

\bibitem{Sho84}
D. Shoenberg, {\it Magnetic oscillations in metals} (Cambridge University Press, Cambridge, 1984).

\bibitem{UPT3}
R. Joynt and L. Taillefer, Rev. Mod. Phys. {\bf 74}, 235 (2002).

\bibitem{OrgMet}
J. Wosnitza {\it et al.}, Phys. Rev. Lett. {\bf 67}, 263 (1991); J. S. Brooks {\it et al.}, {\it ibid.} {\bf
69}, 156 (1992); M. V. Kartsovnik {\it et al.}, {\it ibid.} {\bf 77}, 2530 (1996).

\bibitem{HiTc}
C. M. Fowler {\it et al.}, Phys. Rev. Lett. {\bf 68}, 534 (1992).

\bibitem{2DEG}
J. P. Eisenstein {\ et al.}, Phys. Rev. Lett. {\bf 55}, 875 (1985); J. G. E. Harris {\it et al.}, {\it ibid.}
{\bf 86}, 4644 (2001).

\bibitem{MgB2}
E. A. Yelland {\it et al.}, Phys. Rev. Lett. {\bf 88}, 217002 (2002); A. Carrington {\it et al.}, {\it ibid.}
{\bf 91}, 037003 (2003).

\bibitem{Mn12_Fe8}
R. Sessoli {\it et al.}, Nature {\bf 365}, 141 (1993); L. Thomas {\it et al.}, {\it ibid.} {\bf 383}, 145
(1996); J. R. Friedman {\it et al.}, Phys. Rev. Lett. {\bf 76}, 3830 (1996); C. Sangregorio {\it et al.},
{\it ibid.} {\bf 78}, 4645 (1997).

\bibitem{OW_Mn3x3}
O. Waldmann , L. Zhao, and L. K. Thompson, Phys. Rev. Lett. {\bf 88}, 066401 (2002).

\bibitem{insulator}
Because of the large seperation of individual molecules, Mn-[3~$\times$~3] crystals are good insulators
showing that a dHvA type of effect cannot be responsible for the magneto oscillations observed here.

\bibitem{Zha00}
L. Zhao, C. J. Matthews, L. K. Thompson, and S. L. Heath, Chem. Commun. {\bf 2000}, No. 4, 265 (2000).

\bibitem{Koc03}
R. Koch {\it et al.}, Phys. Rev. B {\bf 67}, 094407 (2003).

\bibitem{Cor00}
A. Cornia {\it et al.}, Chem. Phys. Lett. {\bf 322}, 477, (2000).

\bibitem{FW_osc}
"Oscillations" were presented for e.g. molecular ferric wheels in the {\it field derivative} of the
magnetization and torque, $dm(B)/dB$ and $d\tau(B)/dB$ \cite{FWs,OW_FWs}. They reflect steps in the
magnetization and torque \cite{Cor99,OW_FWs}, and should not be confused with the oscillations observed here.

\bibitem{FWs}
K. L. Taft {\it et al.}, J. Am. Chem. Soc. {\bf 116}, 823 (1994); D. Gatteschi {\it et al.}, Science {\bf
265}, 1054 (1994).

\bibitem{Cor99}
A. Cornia {\it et al.}, Angew. Chem. Int. Ed. {\bf 38}, 2264 (1999).

\bibitem{OW_FWs}
O. Waldmann {\it et al.}, Inorg. Chem. {\bf 38}, 5879 (1999); {\it ibid.} {\bf 40}, 2986 (2001).

\bibitem{Dipdip}
Intramolecular dipole-dipole interactions cannot be neglected, but have similar effects as the single-ion
anisotropy and are well grasped by $S^2_{i,z}$ terms. The $D$ values should be thus understood as to include
both anisotropy contributions.

\bibitem{OW_3x3_NVT}
O. Waldmann (unpublished).

\bibitem{OW_Cr8}
O. Waldmann {\it et al.}, Phys. Rev. Lett. {\bf 91}, 237202 (2003).

\bibitem{Mn3x3_ins}
T. Guidi {\it et al.} (unpublished).

\bibitem{levelmix}
The oscillations were not obtained with the analysis presented in \cite{OW_Mn3x3}, since, following the
accepted approach for systems in the strong exchange limit like Mn-[3~$\times$~3], mixing between different
spin levels was neglected.

\bibitem{Ben90}
A. Bencini and D. Gatteschi, {\it Electron Paramagnetic Resonance of Exchange Coupled Cluters} (Springer,
Berlin, 1990).

\bibitem{OW_MQT_FW}
O. Waldmann, Europhys. Lett. {\bf 60}, 302 (2002).

\bibitem{Sha02}
Y. Shapira and V. Bindilatti, J. Appl. Phys. {\bf 92}, 4155 (2002).

\end{references}
\end{document}